\documentclass[twocolumn,preprintnumbers,floatfix,prb,showpacs,superscriptaddress]{revtex4}
\usepackage{graphicx}
\usepackage{dcolumn} 
\usepackage{bm}
\usepackage{epsfig}
\begin{document} 

\title{Electronic structure of V$_4$O$_7$: charge ordering, metal-insulator transition and magnetism}

\author{A.~S. Botana}
\email{antia.sanchez@usc.es}
\affiliation{Departamento de F\'{i}sica Aplicada,
  Universidade de Santiago de Compostela, E-15782 Campus Sur s/n,
  Santiago de Compostela, Spain}
\affiliation{Instituto de Investigaci\'{o}ns Tecnol\'{o}xicas,
  Universidade de Santiago de Compostela, E-15782 Campus Sur s/n,
  Santiago de Compostela, Spain}
\author{V. Pardo}
\affiliation{Department of Physics,
  University of California, Davis, CA 95616
}
\author{D. Baldomir}
\affiliation{Departamento de F\'{i}sica Aplicada,
  Universidade de Santiago de Compostela, E-15782 Campus Sur s/n,
  Santiago de Compostela, Spain}
\affiliation{Instituto de Investigaci\'{o}ns Tecnol\'{o}xicas,
  Universidade de Santiago de Compostela, E-15782 Campus Sur s/n,
  Santiago de Compostela, Spain} 
\author{A. V. Ushakov}
\affiliation{
II. Physikalisches Institut, Universit\"at zu K\"oln, Z\"ulpicher Str. 77, D-50937 K\"oln, Germany}   
\author{D. I. Khomskii}
\affiliation{
II. Physikalisches Institut, Universit\"at zu K\"oln, Z\"ulpicher Str. 77, D-50937 K\"oln, Germany}  
\email{khomskii@ph2.uni-koeln.de}

\date{\today}

\begin{abstract}

The low and high-temperature phases of V$_4$O$_7$ have been studied by \textit{ab initio} calculations. At high temperature, all V atoms are electronically equivalent and the material is metallic. Charge and orbital ordering, associated with the distortions in the V pseudo-rutile chains, occur below the metal-insulator transition. Orbital ordering in the low-temperature phase, different in V$^{3+}$ and V$^{4+}$ chains, allows to explain the distortion pattern in the insulating phase of V$_4$O$_7$. The in-chain magnetic couplings in the low-temperature phase turn out to be antiferromagnetic, but very different in the various V$^{4+}$ and V$^{3+}$ bonds. The V$^{4+}$ dimers formed below the transition temperature form spin singlets, but V$^{3+}$ ions, despite dimerization, apparently participate in magnetic ordering.


\end{abstract}

\maketitle

\section{Background}

The transition from a localized to an itinerant electron picture has been thoroughly studied in the last decades.\cite{goodenough_mit_book} In particular, 3d electron systems can be described by both pictures, depending on the situation, structure, U/W ratio (on-site Coulomb repulsion vs. bandwidth), pressure, strain, etc.\cite{imada_rmp} Metal-insulator transitions (MIT) have been observed in these compounds induced by pressure,\cite{mit_pressure} chemical pressure,\cite{av2o4_santi,cov2o4_zhou} change in structure,\cite{vo2_mit} temperature,\cite{nickelates_mit} spatial confinement,\cite{svo_sto_ssc} or a simple change in electron count.\cite{la3ni2o7_expt,la3ni2o7_vpardo}

Charge ordering (CO) plays an important role in the insulating side of these MIT in various transition metal oxides, happening both in bulk compounds such as perovskite manganites~\cite{manganites_co} and nickelates~\cite{nickelates_co}, and in nanostructures.\cite{chakhalian_nickelates_films_co} When materials approach a MIT, the lattice can become softer due to bond-length fluctuations,\cite{fran_goody_review} and this could lead to structural distortions or phase transitions of peculiar character.\cite{mgti2o4_khomskii,crn_nmat}

V$_{4}$O$_{7}$ is a member of the homologous series V$_{n}$O$_{2n-1}$= V$_{2}$O$_{3}$+ (n-2)VO$_{2}$ (3$\leq$ n $\leq$9), known as Magn\'eli phase compounds.\cite{magneli} The structure of the members of this series is based on the rutile unit cell. VO$_2$, like several other transition metal dioxides, crystallyzes (above room temperature) in the rutile structure,\cite{vo2_struct} characterized by an octahedral  coordination of the V cations by O atoms and a strong metal-metal bonding along the c-direction. The crystal structure of V$_4$O$_7$ (Ref. \onlinecite{hodeau}) consists of rutile-like blocks that are infinite in two-dimensions and four VO$_6$ octahedra wide in the third (see Fig. \ref{structures}). The shear planes between the rutile blocks have the typical local structure for V$_2$O$_3$ (Ref. \onlinecite{v2o3_struct}). The resulting structure is triclinic, belonging to a $P\bar{1}$ space group, with four and seven independent V and O sites, respectively. The V sites are split into two groups: V3 and V4, which are at the shear plane, and V1 and V2, at the center of the rutile blocks. Since V3 and V4 octahedra are connected by face sharing, their environment is sesquioxide-like whereas that of V1 and V2 is rutile-like. The cations form two independent chains V3-V1-V1-V3 and V4-V2-V2-V4 running parallel to the pseudorutile c-axis (see Fig. \ref{structures}). 


Various members of the series are characterized by a MIT.\cite{vo2_mit} The physics of the transition in VO$_2$ has been strongly debated. This is a correlated compound, but the occurrence of a dimerization along the c-direction, leading to a formation of spin-singlets due to the Peierls distortion, shows that both correlations and Peierls physics play a role in causing the transition.\cite{vo2_peierls1,renata,vo2_corr,vo2_peierls2}
V$_4$O$_7$ shows a MIT at 250 K (T$_c$).\cite{v4o7_mit} The MIT in V$_{4}$O$_{7}$ is a weakly first order one.\cite{marezio_prl,prb_NMR_metallic_insulating_models} In this case, the lattice symmetry is unchanged during the phase transition unlike the cases of V$_{2}$O$_{3}$ and VO$_{2}$. The compound also exhibits a paramagnetic to antiferromagnetic (AFM) transition at T$_N$$\sim$ 40 K.\cite{low_T_afm_nmr} The fact that the MIT occurs at a different temperature than the magnetic one allows to think that the magnetic order is not a driving force of the phase transition. A simple electron count in V$_{4}$O$_{7}$ gives an average valence V$^{3.5+}$ for the metal cations. Hence, 2V$^{3+}$(d$^2$) and 2V$^{4+}$(d$^1$) per formula unit could be expected if CO would occur. Different previous works showed that the metallic state was characterized by an almost complete disorder of the V$^{3+}$ and V$^{4+}$ cations whereas the insulating state showed a high degree of ordering:\cite{marezio_prl,low_T_afm_nmr,prb_NMR_metallic_insulating_models} the V3-V1-V1-V3 chain is formed by V$^{4+}$ ions and the V4-V2-V2-V4 by  V$^{3+}$ ones. To summarize, CO in the cation sites occurs below the transition, and this CO state eventually orders magnetically at lower temperatures. Other works point out that the magnetic moments are localized not only in the insulating phase but to some degree also in the metallic one.\cite{hodeau} Most studies show that, in addition to electron localization, half of the V$^{4+}$ and all V$^{3+}$ cations are paired, forming short V-V bonds along their corresponding chains in the insulating state.\cite{hodeau,prb_NMR_metallic_insulating_models, v4o7_magnetism_neutron,marezio_prl,low_T_afm_nmr} Another feature of the structural changes occuring at the MIT is the ferro- or antiferroelectric-like shift of unpaired V$^{4+}$ ions at the shear planes, from the center of O$_6$ octahedra towards one of the surrounding oxygens. Such distortion (``twisting'') also occurs in VO$_2$.\cite{vo2_peierls1,vo2_dimer,pouget_prb,pouget_prl}

Several questions arise when looking at these results. The first is, what is the nature of V pairing in the insulating phase? Similar pairing in VO$_2$ is largely connected with the orbital repopulation:\cite{haverkort} d-electrons, which in the metallic phase above T$_c$ were more or less equally distributed over all the t$_{2g}$-orbitals, below T$_c$ occupy predominantly one orbital (d$_{xy}$ in the local coordinate system), with the lobes in the c-direction, giving strong overlap with neighboring V's along the chains in the c-axis. The resulting one-dimensional electron system is unstable with respect to Peierls, or spin-Peierls dimerization, leading to the formation of singlet V$^{4+}$-V$^{4+}$ dimers. There exist many other compounds with transition metal ions with configuration d$^1$ (containing V$^{4+}$, Ti$^{3+}$, or Nb$^{4+}$) which also form singlet dimers at low temperatures. Thus, we could think that a similar phenomenon could exist also in V$_4$O$_7$, at least for the V$^{4+}$ chains V3-V1-V1-V3. But, in fact, in these segments only the two middle V's (V1-V1) form singlet dimers, with V3 ions remaining unpaired and apparently magnetic. The question is why is this so.

On the other hand, all the V$^{3+}$ in V$_4$O$_7$ form dimers, which is not very typical for V$^{3+}$ cations (d$^2$, S=1): only very few compounds with V$^{3+}$ form  dimers. This happens for example in ZnV$_2$O$_4$,\cite{znvo_prl} but in this case the short V-V dimers are not in a singlet state, but are ``ferromagnetic'' (FM), i.e. have parallel spins on two V's. What is the character of V$^{3+}$ dimers in V$_4$O$_7$, and why these dimers are formed, is an open question.

Another problem is the character and the mechanism of magnetic ordering in V$_4$O$_7$ below T$_N$ = 40 K. Despite long history of study of V$_4$O$_7$, its magnetic structure is actually not known. It is also not clear which V ions participate in the magnetic ordering. Apparently, the paired V$^{4+}$ ions (V1-V1) are in a singlet state and ``drop out of the game''. If V$^{3+}$ pairs would also be singlet, then only 1/4 of all the V's, the unpaired V$^{4+}$ in V3 position, would be magnetic. But then it would be difficult to understand why such a ``dilute'' magnetic system, with magnetic ions with small spin S= 1/2 and located far from each other, would give a reasonably high value of the N\'eel temperature of $\sim$ 40 K. Alternatively, it is possible that the spins of the V$^{3+}$ ions are not completely quenched, and that these ions also contribute to magnetic ordering.

In order to shed light on the issues of charge and magnetic order and on the nature of the observed structural distortions, to analyze what degree of charge disproportionation exists in both phases and also what is the nature and strength of each of the magnetic couplings, we have studied the electronic structure of both metallic and insulating phases of V$_{4}$O$_{7}$ by means of \textit{ab initio} calculations.\cite{footnote}  
Besides CO, we have obtained the specific orbital ordering (OO) in the insulating phase, which allows to explain the distortion pattern observed in it. We have also analyzed the plausible magnetic configurations and have discussed the magnetic properties of both low and high-temperature phases.


\section{Computational details}

Our electronic structure calculations have been performed with the {\sc WIEN2k} code,\cite{wien2k,wien} based on density functional theory\cite{dft,dft_2} (DFT) utilizing the augmented plane wave plus local orbitals method (APW+lo).\cite{sjo} We have used the LDA+U \cite{sic} approach including self-interaction corrections in the so-called ``fully localized limit'' with the on-site Coulomb repulsion U= 4.1 eV and the on-site Hund's rule coupling J= 0.7 eV. This method improves over the generalized gradient approximation\cite{gga} (GGA) or local density approximation (LDA)\cite{lda} in the study of systems containing correlated electrons by introducing the on-site Coulomb repulsion U. Results presented here are consistent for values of U in the interval from about 3 to 5 eV, a reasonable range according to previous similar calculations of V$^{4+}$ compounds\cite{tio2_vo2_prl,vo2_tio2_mit,srvo3_srtio3_mit} and also of V$^{3+}$ compounds.\cite{znvo_prl} The calculations were fully converged with respect to the k-mesh and R$_{mt}$K$_{max}$. Values used for the k-mesh were 5$\times$4$\times$4 sampling of the full Brillouin zone. R$_{mt}$K$_{max}$= 6.0 was chosen for all the calculations. Selected muffin tin radii were the following: 1.72 a.u. 
for V, and 1.53 a.u. for O in the low-T structure and in the high-T phase, 1.77 a.u. 
for V, and 1.57 a.u. for O.

\section{Discussion of the results}

\subsection{Structure}

\begin{figure}
\begin{center}
\includegraphics[width=\columnwidth,draft=false]{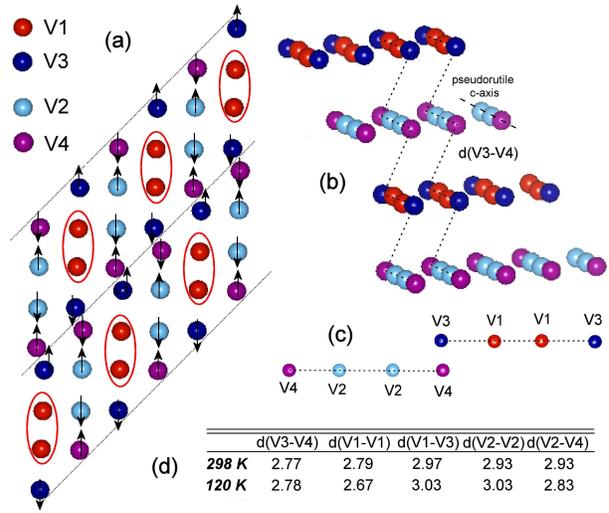}
\caption{(Color online) (a) Pseudorutile blocks in V$_4$O$_7$. The lines represent the shear planes. We also show the magnetic structure of the low-temperature phase obtained for U= 4.1 eV (the coupling accross the shear planes becomes AFM for U $\geq5.4$ eV). The ovals around V1 pairs indicate spin-singlet states. (b) View of the chains where the zig-zag of the structure can be seen. (c) Unit cell of V$_4$O$_7$. (d) Table with the different V-V distances (in \AA) of the high-temperature (metallic) and low-temperature (insulating) structures at 298 K and 120 K, respectively as obtained in Ref. \onlinecite{hodeau}.}\label{structures}
\end{center}                                                                                                                                           \end{figure}

We have taken the structural data for both the metallic phase (at 298 K) and the insulating one (at 120 K) from the work of Hodeau and Marezio.\cite{hodeau} The main structural features are depicted in Fig. \ref{structures}.
The two phases comprise different V-V and V-O distances (the V-V distances are shown in Fig. \ref{structures}(d), the V-O distances are presented in Table \ref{distancias_v_o}). As can be seen, in the V4-V2-V2-V4 chain of the metallic phase, the cations are equidistant. In the insulating state, the V2-V4 distance becomes shorter than the V2-V2 one, favoring a possible V2-V4 pairing. Even larger change occurs in the V3-V1-V1-V3 chain distances from the high to the low temperature structure, with the V1-V1 distance  becoming much shorter at lower temperatures. The distance between chains remains almost unchanged across the MIT.

\begin{table}[h!]
\caption{Interatomic distances between V and O atoms (in \AA) in V$_4$O$_7$ for both the low-temperature (120 K) and high-temperature (298 K) phases. Structural data are taken from Ref. \onlinecite{hodeau}.}\label{distancias_v_o}
\begin{center}
\begin{tabular}{c c c c c c c c c c c}
\hline
\hline
  &&  120 K&& 298 K&& && 120 K&& 298 K     \\
\hline
d(V1-O2) &&  1.81 && 1.88 && d(V3-O3) && 1.73 && 1.78 \\
d(V1-O1) &&  1.88 && 1.89 && d(V3-O6) && 1.92 && 1.94 \\
d(V1-O1) &&  1.92 && 1.93 && d(V3-O5) && 1.95 && 1.96 \\
d(V1-O6) &&  1.94 && 2.05 && d(V3-O7) &&  1.99 && 1.98 \\
d(V1-O4) &&  2.05 && 2.02 && d(V3-O4) &&  2.01 && 2.03\\
d(V1-O5) &&  2.06 && 2.03 && d(V3-O5) &&  2.15 && 2.12 \\
d(V2-O1) &&  1.91 && 1.90 && d(V4-O6) &&  1.90 && 1.81 \\
d(V2-O2) &&  1.94 && 1.95 && d(V4-O5) &&  1.96 && 1.97 \\
d(V2-O4) &&  2.00 && 2.03 && d(V4-O7) &&  1.98 && 1.97 \\
d(V2-O7) &&  2.00 && 2.02 && d(V4-O2) &&  2.00 && 1.94 \\
d(V2-O3) &&  2.03 && 1.97 && d(V4-O7) &&  2.08 && 2.12 \\
d(V2-O3) &&  2.05 && 2.01 && d(V4-O4) &&  2.13 && 2.10 \\

\hline
\end{tabular}
\end{center} 
\end{table}

According to the analysis performed in Ref. \onlinecite{hodeau} by doing bond-length summations with the experimental structural data, the high temperature metallic state is not comprised of completely disordered V$^{3+}$, V$^{4+}$ cations. The 4224 chains contain around a 58$\%$ of the V$^{3+}$ cations whereas the 3113 chains contain the same amount of the V$^{4+}$ ones. Hence, in each chain, even in the metallic state, there remains 
some charge segregation, but a rather weak one.
In the insulating phase, the 4224 chains become richer in 3+ cations (about 83$\%$), V2 having a valence of 3.24+, and V4 a valence of 3.09+. Meanwhile, the 3113 chain has 83$\%$ of the V$^{4+}$ cations, each of them having a valence of 3.83+. Thus, the low-temperature insulating state is characterized by the ordering of the V$^{3+}$ and V$^{4+}$ cations over the four cation sites and by the pairing of half of the V$^{4+}$ (those in V1 positions), and of all the V$^{3+}$ in the short V-V bonds along the corresponding chains. The remaining unpaired V$^{4+}$ ions (V3), on the other hand, shift from the centers of the corresponding O$_6$ octahedra, forming one very short V--O bond (V3--O3) of less than 1.8 \AA, as can be seen in Table \ref{distancias_v_o}.


\subsection{Electronic structure}

\subsubsection*{{\bf B1. Low temperature phase}}

\begin{figure}
\begin{center}
\includegraphics[width=\columnwidth,draft=false]{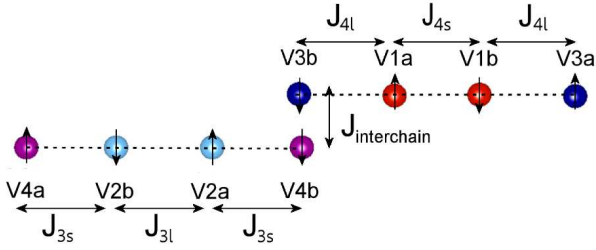}
\caption{(Color online) Magnetic couplings in the unit cell for the low-temperature structure. The AFM in-chain couplings and the inter-chain FM one (stable at lower values of U but very small in value) can be seen.}\label{magnetic_gs}
\end{center}
\end{figure}

The electronic structure of the low-temperature phase of V$_4$O$_7$ can be roughly understood by analyzing the short-long V-V bonds alternating in the structure. These, in particular, modify the magnetic couplings, and also the distinct charge and orbital states for the V cations that such a structure might lead to.
In order to analyze the electronic structure of the compound, we have performed LDA+U calculations for several values of U (3.4 - 5.4 eV). For all the U values, the most stable solutions have in-chain antiferromagnetic (AFM) couplings. By increasing the U value, the inter-chain coupling, even though always almost negligible in value, switches from FM for lower U values to AFM for U $\geq5.4$ eV. The magnetic ground-state (consistent with the in-chain couplings for U ranging from 3.4 to 5.4 eV) can be seen in Figs.~\ref{structures}(a) and ~\ref{magnetic_gs}. Table~\ref{momentos} shows the magnetic moments of each V cation in the low-temperature structure obtained for U= 4.1 eV. From the charge/moment analysis, CO looks substantial. In the V$^{4+}$(3113) chains, the V magnetic moments inside the muffin-tin spheres are about 0.8 $\mu_B$, whereas for the V$^{3+}$(4224) chains the moments are of about 1.4 $\mu_B$, slightly smaller than twice the value of V$^{4+}$, that would be the case in a fully CO state. No noticeable difference in moment is observed between V2 and V4, in contrast to the result of bond length summations that predict a slight difference in valence. The point charge model works very well, but hybridization also plays a role in determining the charge states and magnetic moments.

\begin{table}[h!]
\caption{Projection of the spin magnetic moments of V atoms inside the muffin-tin spheres in the V$_4$O$_7$ magnetic ground-state for the low temperature AFM insulating phase and for the high temperature FM metallic one.}\label{momentos}
\begin{center}
\begin{tabular}{c c c c c}
\hline
\hline
\multicolumn{5}{c}{Low-T phase\hspace{2.3cm}High-T phase}\\ \cline{1-2} \cline{4-5}
Atom  &  Magnetic Moment &&  Atom  &  Magnetic Moment  \\
\hline
V1a(b) &  0.8(-0.8)$~\mu_B$ &&  V1 &  1.3$~\mu_B$ \\
V2a(b) &  1.4(-1.4)$~\mu_B$ && V2 &  1.4$~\mu_B$\\
V3a(b) &  0.8(-0.8)$~\mu_B$ && V3 &  1.2$~\mu_B$\\
V4a(b) &  1.4(-1.4)$~\mu_B$ && V4 &  1.2$~\mu_B$\\
\hline
\end{tabular}
\end{center} 
\end{table}

\begin{figure*}
\includegraphics[width=16.5cm,draft=false]{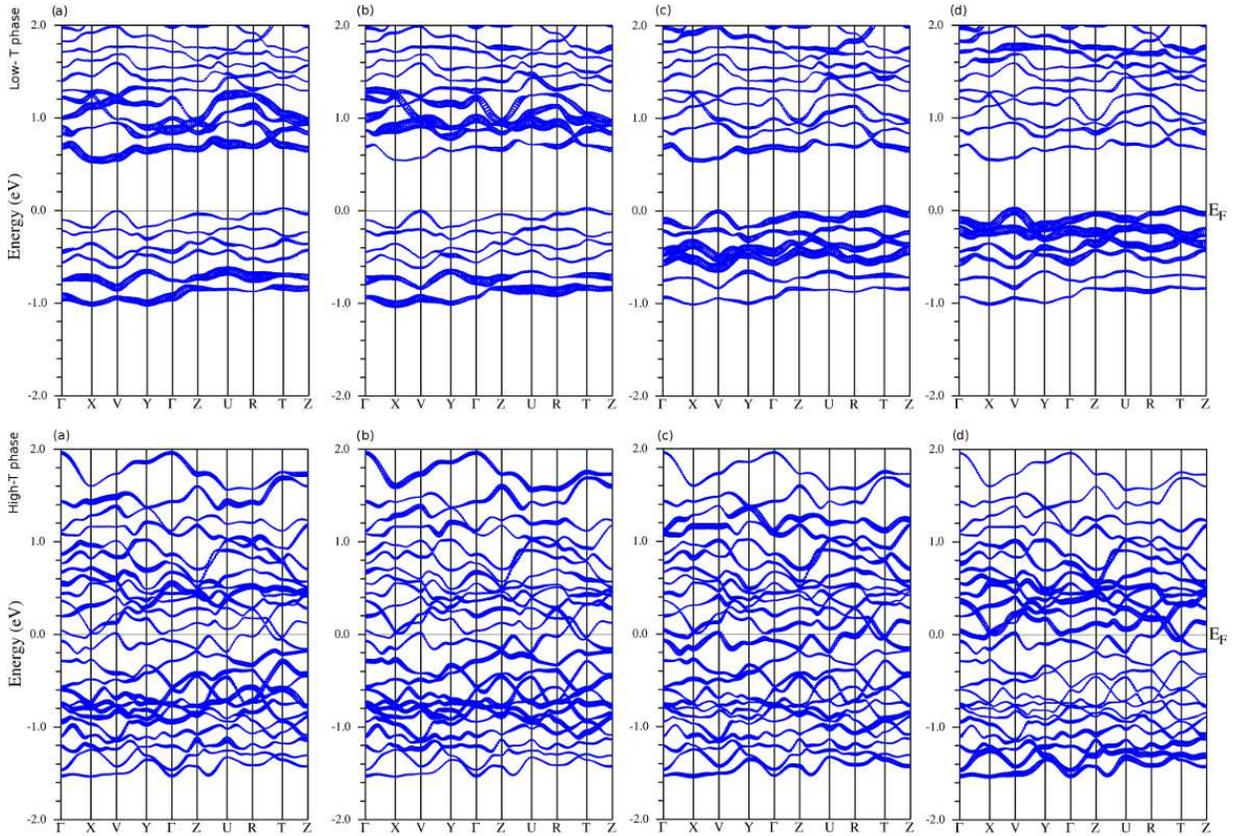}
\caption{(Color online) The upper panels show the band-structure plots with band-character of the majority states for the V cations in the low-temperature insulating phase. (a) Represents the bandstructure plot of the spin-up(down) channel of V1a(V1b), (b) of the spin-up(down) channel of V3a(V3b), (c) of the spin-up(down) channel of V2a(V2b) and (d) of the spin-up(down) channel of V4a(V4b). The differences in the band-structure plots for the V$^{3+}$ (V2 and V4) and the V$^{4+}$ (V1 and V3) cations can be observed. The lower panels show the same plots for the majority (up) states for the V cations in the high-temperature metallic phase (a) for V1, (b) for V2, (c) for V3, and (d) for V4.}\label{bandas}
\end{figure*}

Further indications of the large degree of CO, very close to a d$^1$/d$^2$ state in the corresponding chains, come from the study of the band-structure plots with band-character (the so called ``fat-bands" plots) shown in the upper panels of Fig. \ref{bandas}. A band structure plot is often a more reliable tool to analyze charge states in localized electron systems than a simple electron count inside the muffin-tin spheres.\cite{la2vcuo6} The total charge in both cases would be very similar, causing the very small difference in total energy observed experimentally, but the electronic structure and magnetic moments can be very different. This phase is an insulating one with a d-d gap of about 0.6 eV, for this particular value of U (4.1 eV). About 1 eV below the Fermi level, only d-bands are observed, with the O states much lower (starting at about -2.5 eV), as occurs in other vanadates like VO$_2$,\cite{vo2_tio2_mit} SrVO$_3$,\cite{srvo3_srtio3_mit} sympliflying the picture. We observe, on the upper left two panels of Fig. \ref{bandas}, two cations with a d$^1$ configuration (these are the cations in the V$^{4+}$-chain) and, on the right two panels, other two cations with a distinct d$^{2}$ electronic configuration (those at the V$^{3+}$ chains). The ``fat-bands" plots clearly show one band occupied for the V$^{4+}$ cations, and two for the V$^{3+}$ ones. Close above the Fermi level, for each spin channel, six unoccupied bands can be observed, four of them from the V$^{4+}$: d$^1$ cations and, split, two from the V$^{3+}$: d$^2$ cations.

\begin{figure}
\begin{center}
\includegraphics[width=\columnwidth,draft=false]{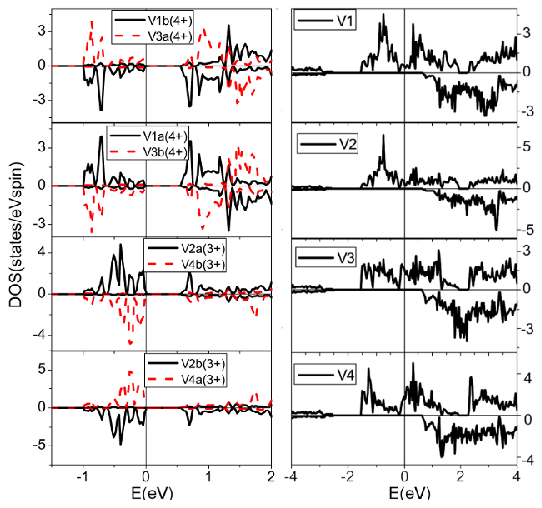}
\caption{(Color online) Left panels represent the partial DOS for each V cation in the low-T phase, where a charge disproportionation effect is clear. Right panels represent that of the V cations in the high-T structure; the different environment leads to a different DOS in each site, but the electron count is similar for all of them. Positive (negative) values indicate up-spin (down-spin) DOS.}\label{dos}
\end{center}
\end{figure}

We can also describe the electronic structure of the material based on the partial density of states (DOS) plots of the various V cations in the low-T structure (see left panels of Fig. \ref{dos}). The different DOS plots for V$^{4+}$ and V$^{3+}$ cations can also be observed, with a different position of the occupied bands with respect to the Fermi level: the relatively narrow V$^{4+}$ bands lie lower in energy, and the broader and more occupied bands corresponding to the V$^{3+}$ cations are closer to the Fermi level. The DOS integrations inside the muffin-tin spheres give a less distinct picture of the CO phenomenon than the band structures, which are much clearer.

\begin{figure}
\includegraphics[width=\columnwidth,draft=false]{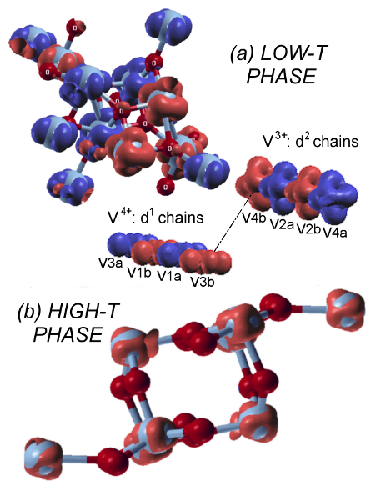}
\caption{(Color online) Three-dimensional plot of the spin density of the magnetic ground-states for both high-T and low-T phases of V$_4$O$_7$, isosurface at 0.1 e/\AA$^3$ obtained using XCrysDen [\onlinecite{xcrysden}]. (a) Represents the plot for the low-temperature phase. Different colors (grayscales) represent spin-up(down) density. The upper panel in (a) shows the density plot for V and O atoms in the unit cell. The lower panel shows the CO character (d$^1$/d$^2$) along each chain in the structure (neither V nor O atoms have been depicted for simplicity). (b) Represents the spin-density plot for the high-temperature metallic phase. All V atoms exhibit a very similar spin density.}\label{rho}
\end{figure}

In order to give further evidence of the large CO character of the low-temperature phase, we can consider a three-dimensional representation of the spin density; this also shows which orbitals are occupied at each site, i.e. it allows to find the OO. In Fig.~\ref{rho}(a) such representation is shown for U= 4.1 eV.

Quite different spin densities (those of V$^{3+}$/V$^{4+}$ cations) along the corresponding chains can be seen. The spin density of the V$^{3+}$(d$^2$) cations corresponds to that of a mixture of d$_{xz}$ and d$_{yz}$ orbitals, whereas for the V$^{4+}$(d$^1$) it can be interpreted as a d$_{xy}$ orbital, using a local coordinate system with the z-axis directed from V towards its apical oxygen.

 In principle, one would expect the local octahedral environments of the V$^{4+}$ and V$^{3+}$ cations to be distorted accordingly, to accomodate one electron in the d$_{xy}$ singlet or two in the d$_{xz}$, d$_{yz}$ doublet. However, the distances we have summarized in Table \ref{distancias_v_o} suggest complicated distortions of the VO$_6$ octahedra, which make it difficult to predict on the basis of only local distortions which orbitals would be occupied,\cite{wu_2011,artem_2011} without analyzing the results of the calculations.

On the basis of these results, one can suggest a possible explanation of the origin of the structural distortions found in V$_4$O$_7$ below T$_c$. V$^{4+}$ ions occupy two types of positions: V1 in the middle of rutile-like chains, and V3 at the shear plane. When one electron occupies the d$_{xy}$ orbital in V$^{4+}$ atoms, which occurs below T$_c$, there exists strong direct overlap of these orbitals along the chain, which can give rise to a metal-metal bonding of V1 ions, like that in the singlet dimers in the insulating phase of VO$_2$.\cite{vo2_tio2_mit} However, the V$^{4+}$ ions in V3 positions are in a different situation: they have a common face with V4 across the shear plane. In this situation, as argued by Goodenough,\cite{vo2_peierls1} the main instability of these ions would be the shift of such V$^{4+}$ from the center of its O$_6$ octahedron (ferro- or antiferroelectric distortion). Indeed, as one sees from the Table \ref{distancias_v_o}, just for this V there exists one very short V-O distance, V3-O3, of 1.73 \AA; all the other V-O distances are larger and more or less equal. Thus, in the structure of V$_4$O$_7$, with shear planes every four sites and with the obtained charge and OO, we would have an (anti)ferroelectric-type displacement of the ``edge'' V3 ions, but the ``inner'' V$^{4+}$ ions V1 would form  short, presumably singlet (see below) dimers. The situation here resembles somewhat  the M2 phase of VO$_2$,\cite{vo2_dimer,pouget_prb,pouget_prl} in which also half of V$^{4+}$ ions form singlet dimers, but the other half has no dimerization, but has instead an antiferroelectric-type distortion (``twisting'') with the formation of one short V-O bond.

\begin{figure}
\begin{center}
\includegraphics[width=\columnwidth,draft=false]{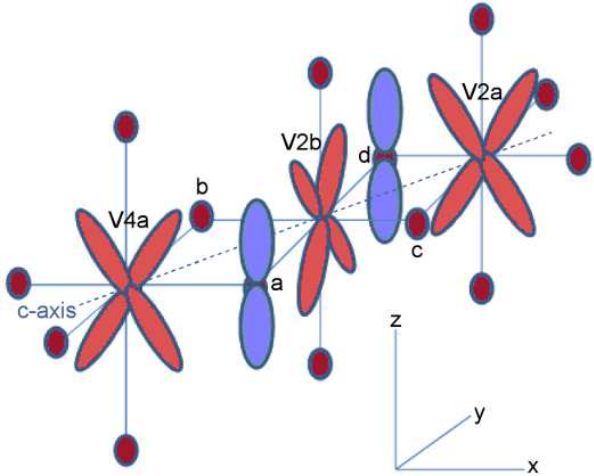}
\caption{(Color online) Schematic representation of the possible hopping processes along the V$^{3+}$ chains in the c-direction. Due to the d$^2$ electronic structure of these cations, the magnetic interactions are mediated by bridging O p-orbitals, contrary to the direct V-V bond (not shown) that occurs in the V$^{4+}$-V$^{4+}$ chains.}\label{orbitals}
\end{center}
\end{figure}

The OO in V$^{3+}$ ions, obtained in our calculations, is, however, quite different from that in V$^{4+}$ cations: in V$^{3+}$ not d$_{xy}$, but rather two other orbitals, d$_{xz}$ and d$_{yz}$,  are occupied by two electrons in the low-temperature phase. These electrons do not have strong direct d-d overlap, but they can instead hop via bridging oxygens. By that process, the d$_{xz}$ electron from one site hops to a d$_{yz}$ orbital of the neighboring V, as sketched in Fig. \ref{orbitals}, where we show the V$^{3+}$ chain running along the c-direction in a more convenient projection. One sees indeed that e.g. the $xz$-electron from the site V4a can hop to a $yz$-state of the site V2b via a p$_z$ orbital of the oxygen $a$, and then again to $xz$-orbital of the site V2a via the p$_z$ orbital of oxygen d. Similarly, there will exist hopping 
($yz$)$_{V4a}$ - (p$_z$)$_b$ - ($xz$)$_{V2b}$ - (p$_z$)$_c$ - ($yz$)$_{V2a}$ \ldots.  As a result, we would have two degenerate half-filled one-dimensional (1D) bands (containing two electrons per V$^{3+}$). Such 1D bands would naturally give rise to Peierls dimerization, which would open the gap in the spectrum and would decrease the energy.\cite{mgti2o4_khomskii} This is a possible explanation of the formation of dimers (V4-V2) and (V2-V4) in V$_4$O$_7$. Similar dimerization was recently found in the rutile-like V$^{3+}$ chains in vanadium hollandite K$_2$V$_8$O$_{16}$,\cite{komarek} apparently due to the same mechanism.


\subsubsection*{{\bf B2. Magnetic order}}

To analyze the possibilility and the consequences of pairing of V atoms along each chain and to understand the magnetic 
properties of the compound, we have calculated the strength of the magnetic couplings in the unit cell. Two kinds of couplings 
can be considered in each chain (one short (s) and one long (l)) plus the inter-chain coupling (see Fig. \ref{magnetic_gs}). 
We have used our total energy calculations for different collinear magnetic configurations to quantify and describe these couplings. 
To that end, we have fitted the total energies by a Heisenberg model, in the form $H= \frac{1}{2}\sum^{}_{i,j}J_{ij}S_iS_j$, with S=1 for 
the V$^{3+}$ sites and S=1/2 for the V$^{4+}$ ones. The V$^{4+}$(d$^1$) sites would be expected to undergo singlet spin pairing whereas 
the V$^{3+}$(d$^2$) ones could pair in singlet, triplet or quintuplet spin states. The values of the in-chain coupling constants, obtained 
in our calculations, are: J$_{3l}\approx$ 40 K, J$_{3s}\approx$ 100 K, J$_{4l}\approx$ 10 K and J$_{4s}\approx$ 1000 K (3 and 4 stand 
for V$^{3+}$ and V$^{4+}$ chains, respectively, and ``l/s" stand for the long and short bonds).

The short-bond magnetic coupling  between paired V$^{4+}$ cations (J$_{4s}$) is very large, $\approx 1000$ 
K,  indicating that most probably these ions form spin-singlet states, reminiscent of the ordering along the c-axis that happens in VO$_2$.\cite{vo2_peierls2} The situation is different in V$^{3+}$ chains, where also the short-bond magnetic coupling (J$_{3s}$) is larger than that in the long bond (J$_{3l}$), but the difference is not as drastic as in the V$^{4+}$ chains. This suggests that spin singlets would not be formed in the V$^{3+}$ chains. The magnetic coupling strength between chains is quite small, but apparently sufficient to form long-range magnetic ordering below T$_{N}=40$ K. We can thus propose that the magnetic structure of this state may be the AFM ordering in V$^{3+}$ segments, coupled to V$^{4+}$ ions V3 with unpaired spins (see Figs. \ref{magnetic_gs} and \ref{structures}(a)). 
Due to the very small inter-chain magnetic coupling, two V$^{4+}$ cations per chain (V3) could be left partially disordered, even at low temperatures.\cite{footnote_2}


Previous works\cite{prb_NMR_metallic_insulating_models,v4o7_magnetism_neutron} give further support to the importance of the pairing interactions as one of the driving forces for the MIT. In the work by Gossard \textit{et al.},\cite{prb_NMR_metallic_insulating_models} the pairing in the V$^{4+}$ chain (corresponding to our J$_{4s}$) becomes complete only below 150 K (100 K below the MIT transition). The magnetic coupling J$_{4s}$ estimated using their nuclear-resonance frequency shifts is also large (J$\sim$ 500 K at 100 K). The nuclear magnetic resonance (NMR) spectrum of spin-echo amplitudes observed at 4.2 K implies a continuous singlet spin pairing in the AFM state, with the  singlet V$^{4+}$ spin-pairs unbroken by the AFM ordering. This is in agreement with our picture. 

In addition, the possible formation of singlet spin pairs along the V$^{3+}$ chains is also discussed in Ref. \onlinecite{prb_NMR_metallic_insulating_models}. Due to the two echo lines appearing in the NMR spectrum one possibility is that i) one line arises from unpaired V$^{4+}$ with a moment of 1$~\mu_B$, while the other comes from paired V$^{3+}$ atoms which have reduced moments of 1$~\mu_B$, due to covalent singlet bonding of one of their two 3d electrons. The other option is: ii) both lines arise from unpaired 4+ atoms that may have become crystallographically inequivalent in the AFM state. Structural results\cite{hodeau} at 120 K, above the magnetic order temperature (T$_N$), indicate the unpaired V$^{4+}$ at the shear planes are not inequivalent. Heidemann \textit{et al.}\cite{v4o7_magnetism_neutron} gave futher confirmation of the  option i) and hence of the partial singlet spin pairing in the V$^{3+}$ chains, possibly with some magnetism thereof remaining.

From our calculations, we can argue that although  J$_{3s}$ is larger than the coupling of V$^{3+}$ spins in the long bond, we do not expect real spin-singlets to be formed in the V$^{3+}$ chains, but a more complicated spin structure is expected to develop. From our calculations, we can see that the magnetic moments obtained for the V$^{3+}$(d$^2$) atoms are reduced by hybridizations from the purely ionic picture. Also, the magnetic moment of these cations barely changes across the transition, as we saw in Table \ref{momentos}.
Thus we propose that both V$^{3+}$ and half of V$^{4+}$ ions (those in V3 positions) participate in long-range magnetic ordering (see Fig.~\ref{structures}(a)).

\subsubsection*{{\bf B3. High temperature phase}}

The electronic structure of the high temperature/metallic phase does not present the large charge-ordering character, as the bond-length summations correctly suggest.\cite{hodeau} We have performed LDA+U calculations for the same magnetic configurations used for the low-temperature structure with the same U values. The magnetic ground-state is now FM and all the sites are almost equivalent: the electronic structure is close to V$^{3.5+}$ for all the atoms in the structure. The magnetic moments of the ground-state solution that can be seen in Table \ref{momentos}, are consistent with this picture. Some difference is seen in the magnetic moments due to the different bandwidths of the respective d-bands,  caused by the different environments the V cations are in, but no trace of CO remains at high temperature, leading to a metallic solution. Also, by looking at the band structures with ``fat-bands" (lower panels of Fig. \ref{bandas}), we do not see the clear d$^{1}$/d$^{2}$ ionic picture we observe in the insulating low-temperature phase. The FM high-temperature band structure is consistent with the absence of charge-ordering we are describing, since it shows more evenly distributed occupations in different bands. The spin density plot for the high temperature phase (see Fig. \ref{rho}(b)) gives further evidence of this absence, confirming a noninteger charge occupation (roughly V$^{3.5+}$) at all the V sites. Even though a different DOS is seen for each of the V sites in the high temperature structure (right panels of Fig. \ref{dos}), the electron count is very similar for each of them, as the ``fat bands" and magnetic moments also show. The FM ordering (or paramagnetic state experimentally found at high temperatures) also increases the bandwidths when compared to the low-temperature CO AFM phase. The absence of CO is supported by experimental works.\cite{hodeau,prb_NMR_metallic_insulating_models,v4o7_magnetism_neutron} The high temperature susceptibility data for V$_4$O$_7$\cite{prb_NMR_metallic_insulating_models,v4o7_magnetism_neutron} are not conclusive about the sign of the Curie-Weiss temperature due to the high transition temperature. Our calculations suggest a change to a FM nature of the in-chain couplings in the high-temperature phase. 
One can qualitatively interpret the FM coupling in the high-temperature metallic phase as a double exchange interaction.

\subsection{Final remarks}

Thus we see that, in analogy with VO$_2$, in the V$^{4+}$ chains there occurs an OO facilitating singlet pair formation. Also, the analysis of the electronic structure helps to understand why only the inner (V1-V1) V$^{4+}$ ions form such pairs, whereas (V4-V2) and (V2-V4) pairs are formed in V$^{3+}$ chains. This has to do with the orbital occupation in the V$^{3+}$ chains, different from that in the V$^{4+}$ chains: two orbitals, d$_{xz}$ and d$_{yz}$, are occupied by two electrons at these sites, in contrast to d$_{xy}$ orbital occupation in V$^{4+}$ chains. Apparently, this is responsible for somewhat weaker dimerization (the changes in the V-V distance across the CO transition are, $\Delta$(V4-V2)$_{high-T,low-T}$= 0.1 \AA, as compared with $\Delta$(V1-V1)$_{high-T,low-T}$= 0.12 \AA, see Fig. 1d). This also leads to smaller modulation of the exchange constants within the V$^{3+}$ chains (J$_{short}$ = 100 K vs J$_{long}$ = 40 K), whereas this difference is much larger in V$^{4+}$ segments (J$_{short}$ = 1000 K vs J$_{long}$ = 10 K). As a result, apparently V$^{3+}$ ions do not form singlets, despite dimerization, but participate in magnetic ordering, together with unpaired V$^{4+}$ (V3) ions. This can explain the appearance of magnetic ordering at a relatively high temperature T$_N$= 40 K.

\section{Summary}

In this paper, we have studied the electronic structure of V$_4$O$_7$ analyzing the differences between the low-temperature (charge ordered, antiferromagnetically ordered and insulating) and the high-temperature (charge disordered, paramagnetic and metallic) phases. 

Calculations show the strong CO in the low-temperature phase that, together with structural distortions, lead to a gap opening. The high-temperature structure is charge disordered, with all V atoms in a similar $\sim$~d$^{1.5}$ electronic configuration. All the in-chain magnetic couplings in the low-temperature phase are AFM, and the inter-chain couplings are very weak (and of different signs depending on the U-value). Magnetic coupling changes to FM in the high-temperature phase as a result of the change in electron distribution caused by charge disorder. The AFM couplings in the low-temperature phase are obtained, being very strong in the V$^{4+}$-V$^{4+}$ short bonds, one order of magnitude stronger than the short V$^{3+}$-V$^{3+}$ magnetic couplings (and two orders of magnitude stronger than in the V$^{4+}$-V$^{4+}$ long bonds). This suggests the formation of a spin-singlet between the ``middle'' V's within the V$^{4+}$ chains, leaving V3 atoms with unpaired spins. The couplings in the V$^{3+}$ chains are more complicated, showing signs of slight moment reduction due to hybridizations. 

The OO obtained in our electronic structure calculations (V$^{4+}$:~d$_{xy}$ vs. V$^{3+}$:~d$_{xz}$,d$_{yz}$) also helps to understand the different bond distance alternations that occur in the two different types of V chains: in the V$^{4+}$ chains there is a strong metal-metal bond between the ``middle'' V1 ions forming spin singlets, and an (anti)ferroelectric-type displacement of the ``edge'' V$^{4+}$ ions at the shear planes. On the other hand,  the V$^{3+}$ chains, due to different orbital occupation, form two one-dimensional-like bands due to bridging with oxygens, that affects the type of coupling along these chains and leads, in particular, to a dimerization of V$^{3+}$ chains, which, however, is not strong enough to make the short V$^{3+}$-V$^{3+}$ dimers spin singlets.

As a result, most probably  both the unpaired ``edge'' V$^{4+}$ and all V$^{3+}$ ions contribute to magnetic ordering below T$_N$, the detailed type of which is yet unknown. It would be very interesting to check the proposed magnetic ordering in V$_4$O$_7$ experimentally.


\acknowledgments
The authors thank C.D. Batista for fruitful discussions. We also acknowledge the CESGA (Centro de Supercomputaci\'on de Galicia) for the computing facilities and the Ministerio de Educaci\'{o}n y Ciencia (MEC) for the financial support through the project MAT2009-08165. Authors also thank the Xunta de Galicia for the project INCITE08PXIB236053PR. A.~S. Botana thanks MEC for a FPU grant. The work of D.I. Khomskii and A.V. Ushakov in K\"oln was supported by SFB 608, FOR 1346, and by the European project SOPRANO.


\begin{thebibliography}{53}
\bibitem[{foo({\natexlab{a}})}]{footnote}
\bibinfo{note}{Some previous electronic-structure calculations of
  V$_{4}$O$_{7}$\cite{es_lda,es_lda_2} have been performed within the local
  density approximation\cite{lda} (LDA) method, but they did not include
  correlation effects that are the key to understand the charge ordered
  insulating phase.}
\bibitem[{foo({\natexlab{b}})}]{footnote_2}
\bibinfo{note}{One should note that the one-electron picture of the electronic
  structure provided by DFT calculations cannot account for the formation of
  spin-singlets between two neighboring atoms. One can only study the case of
  neighboring atoms with local spins coupled AFM to one another, but not a
  multielectron state forming a spin-singlet. Usually such approximation gives
  reasonable description of the gross features of such systems, but may be not
  of the fine details, e.g. in spectroscopy\cite{vo2_peierls2}}.
\end{thebibliography}

\end{document}